\documentclass[aps,prl,twocolumn,superscriptaddress]{revtex4-2}

\usepackage[T1]{fontenc}
\usepackage{braket}
\usepackage{graphicx}
\usepackage[dvipsnames]{xcolor}
\usepackage[colorlinks=true, allcolors=NavyBlue]{hyperref}
\usepackage{soul}
\usepackage{mathtools, amsfonts}
\usepackage{physics,siunitx}

\def\texp{t_{\rm exp}}
\def\na{N_A}
\def\nc{N_C}
\def\cA{\mathcal{A}}
\def\cC{\mathcal{C}}
\def\cT{\mathcal{T}}
\def\cF{\mathcal{F}}
\def\pfa{p_{\rm FA}}
\def\fhat{\hat{f}}
\def\thhat{\hat{\theta}}
\def\ahat{\hat{a}}

\begin{document}

\title{Entanglement-Enhanced Quantum Nano-Vibrometry}

\author{Colin P. Lualdi}
\email[Contact author: ]{clualdi2@illinois.edu}
\affiliation{Department of Physics, The Grainger College of Engineering, University of Illinois Urbana-Champaign, Urbana, IL, USA}
\affiliation{Illinois Quantum Information Science and Technology Center, The Grainger College of Engineering, University of Illinois Urbana-Champaign, Urbana, IL, USA}

\author{Joshua Rapp}
\affiliation{Mitsubishi Electric Research Laboratories, Cambridge, Massachusetts 02139, USA}

\author{Spencer J. Johnson}
\affiliation{Department of Physics, The Grainger College of Engineering, University of Illinois Urbana-Champaign, Urbana, IL, USA}
\affiliation{Illinois Quantum Information Science and Technology Center, The Grainger College of Engineering, University of Illinois Urbana-Champaign, Urbana, IL, USA}

\author{Michael Vayninger}
\affiliation{Department of Physics, The Grainger College of Engineering, University of Illinois Urbana-Champaign, Urbana, IL, USA}
\affiliation{Illinois Quantum Information Science and Technology Center, The Grainger College of Engineering, University of Illinois Urbana-Champaign, Urbana, IL, USA}

\author{Paul G. Kwiat}
\affiliation{Department of Physics, The Grainger College of Engineering, University of Illinois Urbana-Champaign, Urbana, IL, USA}
\affiliation{Illinois Quantum Information Science and Technology Center, The Grainger College of Engineering, University of Illinois Urbana-Champaign, Urbana, IL, USA}

\begin{abstract}
The study of dynamic systems at the nanometer scale can benefit from the loss and background resilience offered by quantum two-photon interference. However, fast measurements with the required resolution are difficult to realize. As a solution, we introduce extreme energy entanglement between the photons undergoing interference. Using a flux probing analysis technique, we recover vibrational signals with frequencies as high as 21 kHz. Along with validating nanometer-scale precision and accuracy, we observe a significant quantum advantage when measuring in the presence of loss and background.
\end{abstract}

\maketitle

Quantum two-photon interference \cite{hong_measurement_1987} enables precision sensing by quantifying the indistinguishability of two photons \cite{scott_beyond_2020} incident on a balanced beamsplitter. Diverse applications that leverage the photons' temporal degree of freedom for relative-delay measurements include quantum optical coherence tomography (QOCT) \cite{abouraddy_quantum-optical_2002, nasr_demonstration_2003}, quantum microscopy \cite{ndagano_quantum_2022}, clock synchronization \cite{bahder_clock_2004, xie_implementation_2021}, quantum positioning \cite{bahder_quantum_2004, yang_two-parameter_2019}, and fundamental metrology \cite{dauler_tests_1999, giovannini_spatially_2015, restuccia_photon_2019}. As a fourth-order effect, two-photon interference offers features not found in second-order ``classical'' single-photon interference. In particular, measuring in coincidence provides robustness against imbalanced path loss and optical background \cite{lualdi_fast_2025}. Furthermore, the interference manifests as a ``dip'' in the coincidence rate as a function of relative delay between the two input modes of a beamsplitter. As this Hong-Ou-Mandel dip is typically much wider than the photons' wavelength, the resulting phase insensitivity can relax constraints in many contexts.

This phase insensitivity also complicates achieving high measurement resolution, particularly when probing dynamic systems. While Fisher-information analysis has enabled nanometer-scale (or, equivalently, attosecond-scale) resolution even with a relatively wide dip, a recent demonstration required $O(10^{11})$ photon pairs collected over hours \cite{lyons_attosecond-resolution_2018}. Alternatively, applications such as QOCT often leverage ultra-broadband photons to realize narrow dips. However, despite decades of development, attained QOCT resolutions remain limited to hundreds of nanometers \cite{nasr_submicron_2008, okano_054_2015} and a low pair rate continues to be cited as a significant limitation \cite{dabrowska_quantum-inspired_2024}.

Consequently, optical measurement of time-varying signals typically rely on other techniques, such as laser Doppler vibrometry (LDV) \cite{ rothberg_international_2017}, frequency-modulated continuous-wave lidar \cite{behroozpour_lidar_2017}, intensity measurements \cite{rehain_single-photon_2021}, or time-of-flight (ToF) sensing \cite{kitichotkul_simultaneous_2025}. However, many of these techniques offer complementary capabilities. For example, commercial LDVs offer sub-nanometer resolution but, being classical interferometers, are susceptible to loss. Conversely, ToF methods are constrained by detector timing resolution, with picoseconds representing the state of the art \cite{mccarthy_high-resolution_2025, shahverdi_mode_2018}.

In this Letter, we show how two-photon interference can be enhanced to enable the study of nanometer-scale vibrations in the presence of background and loss. By combining highly non-degenerate energy entanglement with analysis techniques drawn from single-photon imaging, high resolution and bandwidth can be achieved simultaneously. We report accurate reconstruction of vibrational signals at frequencies through at least 21 kHz. Our method thus enables dynamic measurements in settings where loss, background, and photosensitivity can be factors, such as long-range vibrometry or monitoring biological samples. 

\textit{Entanglement-enhanced interference}\textemdash In two-photon interferometry, the per-photon information content may be increased by introducing energy entanglement between the photons undergoing interference. A sinusoidally modulated dip results, with a period dependent on the difference of the entangled energies \cite{ou_observation_1988, rarity_two-color_1990}. Specifically, when two photons in the state
\begin{equation}
    \ket{\psi} = \frac{1}{\sqrt{2}}\left(\ket{\omega_1}_a\ket{\omega_2}_b + \ket{\omega_2}_a\ket{\omega_1}_b\right)
\end{equation}
impinge on a balanced beamsplitter via inputs $a$ and $b$, they exit in separate outputs and result in a coincidence detection with a probability of
\begin{equation}
    P_C = \frac{1}{2}\left\{1-\cos\left[\left(\Delta \omega \right) \tau \right]e^{-2\sigma^2\tau^2} \right\}.
    \label{eq:fringe}
\end{equation}
Here, $\omega_i$ denotes the photon's angular frequency, $\Delta\omega \equiv \omega_1 - \omega_2$ is the angular frequency detuning of the entangled energies, $\tau$ is the relative temporal delay between inputs $a$ and $b$, and $\sigma$ is the photons' angular frequency half bandwidth. 

Multiple experiments \cite{chen_hong-ou-mandel_2019, torre_sub-m_2023} have successfully combined this effect with the Fisher information technique introduced by Ref.~\cite{lyons_attosecond-resolution_2018} to demonstrate measurement resolutions of hundreds of nanometers with only $O(10^4)$ photon pairs; the  speed of light facilitates conversion between temporal and spatial delays. When estimating $\tau$ with this technique, the standard deviation $\sigma_\tau$ is constrained by the the quantum Cram\'er-Rao bound (QCRB),
\begin{equation}
     \sigma_\tau \geq \sigma_{\tau,\text{QCRB}} = \frac{1}{\sqrt{\mathstrut N}}\frac{1}{\sqrt{\mathstrut (\Delta\omega)^2 + 4\sigma^2}}, 
     \label{eq:cramer_rao}
\end{equation}
where $N$ is the number of measurements. In the ideal case (e.g., perfect fringe visibility), the QCRB is saturated \cite{chen_hong-ou-mandel_2019}. Equation \ref{eq:cramer_rao} reveals how the typical limitations set by $N$ and $\sigma$ may be bypassed by increasing $\Delta\omega$.

To improve the resolution by two orders of magnitude and reach the nanometer regime, we leveraged highly non-degenerate polarization entanglement to realize $\Delta\omega = 2 \pi\cdot177$~THz \cite{lualdi_fast_2025} with photon wavelengths of 1550 and 810 nm, up from the tens of THz in Refs. \cite{chen_hong-ou-mandel_2019, torre_sub-m_2023}. Per-photon information content is greatly increased: we observed 1.26 nm (4.2 as) resolution with 59,000(1000) pairs, representing an 88\% saturation of the QCRB. Crucially, as our sensor can achieve \mbox{$>$150,000} detected pairs per second, nanometer-scale resolution is attainable in less than 1 second in the static case. Probing dynamic systems is thus a realistic endeavor. For a detailed experimental and theoretical description of our interferometer, please refer to Ref. \cite{lualdi_fast_2025}.

\begin{figure}[!htb]
\includegraphics{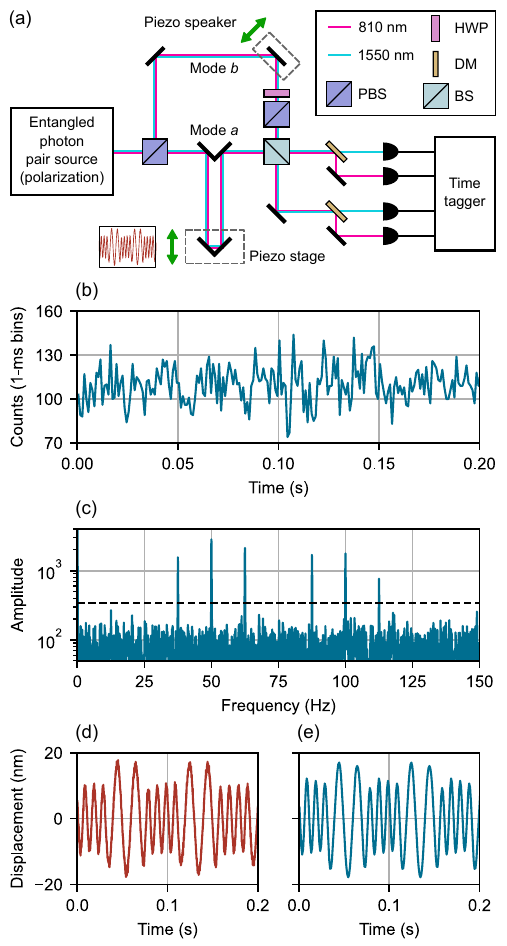}
\caption{\label{fig:fig_1} (a) Apparatus schematic. PBS, polarizing beamsplitter; HWP, half-wave plate; DM, dichroic mirror; BS, 50:50 beamsplitter. (b) Coincidences as a function of time for an input sinusoid with a 25-Hz oscillation between 50 and 100 Hz. The peak-to-peak amplitude also oscillates between $\sim$36~nm and $\sim$22~nm, respectively. The anti-coincidences are not shown. (c) The probed spectrum (2.1 million detected pairs, 10 s). The dashed line indicates the threshold set by $\pfa= 0.1\%$; see text. (d-e) The original and reconstructed signals, respectively. The recovered amplitude is $\sim$35 ($\sim$23) nm for the 50-Hz (100-Hz) portion.}
\end{figure}

\textit{Sensing protocol}\textemdash Figure \hyperref[fig:fig_1]{\ref*{fig:fig_1}(a)} presents the experimental apparatus. A polarizing beamsplitter and half-wave plate converts non-degenerate polarization entanglement into energy entanglement \cite{ramelow_discrete_2009}. Energy-resolved detection measures the coincidence probability $P_C$ by monitoring the rates of both coincidences (photons detected in opposite ports) and anti-coincidences (photons detected in the same port). To sense vibrations, the fringe contrast is first maximized by setting $\tau \approx 0$. Given our non-ideal visibility ($\sim$80--90\%), we then maximize the Fisher information by fine-tuning $\tau$ such that $P_C \approx 0.5$. Finally, vibrations are simulated by varying $\tau$ via either a retro-reflector mounted atop a piezoelectric nano-positioning stage (mode $a$) or a piezoelectric speaker attached to the backside of a mirror (mode $b$).

Photon-pair detections are then monitored as a function of time. Since achieving a target resolution requires a certain number of measurements $N$ (Eq. \ref{eq:cramer_rao}), one could create time bins containing \mbox{$\sim$$N$} counts (Fig.~\hyperref[fig:fig_1]{\ref*{fig:fig_1}(b)}). However, the source brightness then significantly constraints the bin size, which limited recent two-photon interference experiments to Hz-scale sampling rates  \cite{johnson_toward_2023, singh_near-video_2023}.

\begin{figure*}[!htb]
\includegraphics{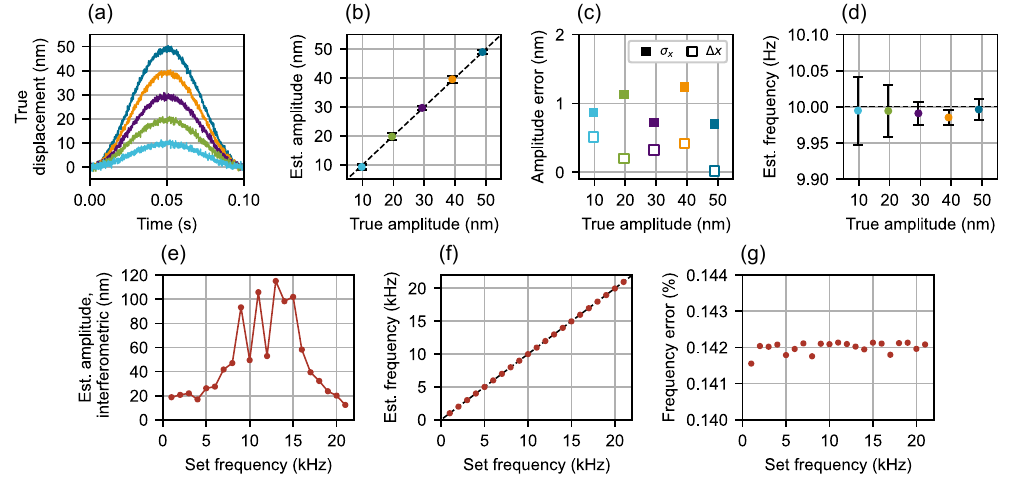}
\caption{\label{fig:fig_2} (a) Signals used for amplitude validation, driving at 10 Hz. (b--d) Estimated mean amplitudes and frequencies (10 trials), along with amplitude precision ($\sigma_x$) and accuracy $(\Delta x \equiv |x_\text{true} - x_\text{measured}|$). The amplitude estimation accounts for the factor of 2 resulting from retro-reflection. Error bars in (b) and (d) show the standard deviation (i.e., precision). Each 1-s trial involved $\sim$190,000 detected pairs. (e--f) Estimated amplitudes and frequencies from a single discrete sine sweep (1--21 kHz in 1-kHz steps). We probe each frequency with $\sim$1 million detected pairs \mbox{(5 s)}. We attribute the consistent $\sim$0.142\% frequency offset error to imperfections in the piezoelectric speaker playback pipeline; a similar error is observed when recording the acoustic signal with a microphone.}
\end{figure*}

To increase the sensing bandwidth, we developed a new analysis protocol based on a ``flux probing'' technique described by Wei et al. \cite{wei_passive_2023}. The detected coincidences and anti-coincidences are each modeled as a realization of a Poisson process~\cite{hayat_theory_1999} with the flux functions
\begin{equation}\label{eq:fluxes}
    \begin{split}
        \varphi_C(t) &= R_C P_C[\tau(t)] \\
        \varphi_A(t) &= R_A \{1 - P_C[\tau(t)]\}.
    \end{split}
\end{equation}
The ratio $R_C/R_A$ can be calibrated from coincidence counts at known delays.
We assume the time-dependent delay $\tau$ corresponds to a vibration described by a small number of sinusoidal components, leading to the model
\begin{equation}
    \varphi(t) = a_0 + \sum_{k=1}^K a_k \cos(2\pi f_k t + \theta_k).
\end{equation}
We define $\texp$ as the duration of a measurement that results in $\nc$  coincidences with timestamps \mbox{$\cC = \{T_i^C\}_{i=1}^{\nc}$} and $\na$ anti-coincidences with timestamps \mbox{$\cA = \{T_i^A\}_{i=1}^{\na}$}. Following Wei et al., we can perform spectral estimation by projecting a set of timestamps $\cT$ onto windowed Fourier basis functions as 
\begin{equation}
    p_f^w(\cT) = \frac{1}{\texp} \sum_{T_i\in\cT} w(T_i) \exp(-\text{i}2\pi f T_i).
\end{equation}
This approach allows Fourier analysis up to a limit set by the timestamp resolution ($\SI{100}{\pico\second}$ in our current setup, corresponding to $\SI{5}{\giga\hertz}$) rather than the bin width of a coincidence count histogram.

We first detect frequencies with significant energy (Fig.~\hyperref[fig:fig_1]{\ref*{fig:fig_1}(c)}) by combining the timestamp sequences as 
\begin{equation}
    y_f^w(\cC, \cA) = p_f^w(\cC) - (\flatfrac{R_C}{R_A}) p_f^w(\cA),
\end{equation}
which is computed for $M$ frequencies $0, \dots, f_{M-1}$ with uniform spacing $\Delta f = \flatfrac{0.6}{\texp}$ using the Hann window 
$w(t) = \cos^2(\flatfrac{\pi t}{\texp})$ to suppress sidelobes.

We retain a set of frequencies $\cF$ where the spectrum magnitude $|y_{f_m}^{w}(\cC, \cA)|$ is greater then a threshold $\kappa$~\cite{rapp_multi-layered_2024}
\begin{equation}
\begin{split}
    \kappa &= \frac{1}{\texp} \sqrt{-\log\{1-[1-\pfa]^{1/M}\}} \\
    & \quad \times \sqrt{\sum_{T_i \in\cC} w^2(T_i) + \left(\flatfrac{R_C}{R_A}\right)^2 \! \sum_{T_i \in\cA} w^2(T_i)},
\end{split}
\end{equation}
where $\pfa$ is the false alarm probability. To avoid the limitations of the initial discretization, we use Brent's method~\cite{brent_algorithms_1973} to estimate a frequency $\fhat_k$ by numerically maximizing $|y_f^w(\cC,\cA)|$ over \mbox{$[f_k - \Delta f, f_k + \Delta f]$} for each $f_k\in\cF$, using a rectangular window $w(t)=1$. The hat denotes an estimated quantity.

For each $\fhat_k$, we estimate the phase as $\thhat_k = \angle y_{\fhat_k}^{w}(\cC, \cA)$.
The sinusoid amplitudes are estimated separately for each timestamp sequence, e.g., for the coincidences, $\ahat_0^C = N_C/\texp$ and 
\begin{equation}
    \ahat_k^C = \frac{2}{\texp} \sum_{T_i\in\cC} \cos(2 \pi \fhat_k T_i + \thhat_k), \quad k \geq 1.
\end{equation}
After reconstructing the flux functions with these estimates, we obtain the coincidence probability as
\begin{equation}
    \hat{P}_C = \frac{\hat{\varphi}_C(t)}{\hat{\varphi}_C(t) + (R_C/R_A)\hat{\varphi}_A(t)}.
\end{equation}
We then recover the original signal via 
\begin{equation}
    \hat{\tau}(t) = (\Delta\omega)^{-1}\cos^{-1}\!\left[(1 - 2 \hat{P}_C)/V_0 \right],
\end{equation}
where $V_0$ is the pre-calibrated fringe visibility. The original and recovered signals are shown in Figs.~\hyperref[fig:fig_1]{\ref*{fig:fig_1}(d)} and \hyperref[fig:fig_1]{\ref*{fig:fig_1}(e)}, respectively. 

\textit{Experimental results}\textemdash Figure~\ref{fig:fig_2} presents a systematic evaluation of amplitude and frequency estimation. To validate amplitude estimation, we leverage the position encoder in the piezoelectric nano-positioning stage to introduce vibrations with independently verifiable amplitudes. Fig.~\hyperref[fig:fig_2]{\ref*{fig:fig_2}(a)} shows the test signals as recorded by the stage: five 10-Hz sinusoidals with peak-to-peak amplitudes ranging between 10 and 50 nm in 10-nm increments. Ten 1-second measurements were taken for each amplitude and the error estimated via the standard deviation. As Figs.~\hyperref[fig:fig_2]{\ref*{fig:fig_2}(b)-(d)} show, we observe nanometer-scale precision and accuracy for amplitude estimation, accompanied by centi-hertz frequency uncertainty.

To investigate frequency estimation, we instead utilize the piezoelectric speaker since our nano-positioning stage is limited to frequencies of a few hundred Hz by its resonant frequency and load mass. While the speaker lacks an integrated position encoder, it can access frequencies in the tens of kHz regime. We validate our frequency recovery by performing a discrete sine sweep from 1 kHz through 21 kHz (1-kHz increments) with the speaker volume fixed at an arbitrary setting compatible with the dynamic range of our interferometer. Each frequency is probed for 5 seconds and analyzed individually. Fig.~\hyperref[fig:fig_2]{\ref*{fig:fig_2}(e)} shows the estimated interferometric peak-to-peak amplitude (i.e., the displacement detected by the interferometer). The observed peaks are attributed to resonances in the speaker-mirror system. The corresponding estimated frequencies and their deviations from the set values are shown in Figs.~\hyperref[fig:fig_2]{\ref*{fig:fig_2}(f--g)}. Since the observed $\sim$0.14\% offset error was similarly present in microphone recordings, we attribute this error to our audio playback pipeline, rather than our interferometric measurement.    

We next demonstrate the practical quantum advantage offered by two-photon interference. Interferometric sensing can benefit from fringes that are unaffected by measurement conditions, such as imbalanced path loss or optical background. For example, many analysis protocols (including the one described in this Letter) rely on a known fringe visibility. Maintaining high visibility also preserves high measurement resolution. The fringes produced by our quantum interferometer are thus advantageous as they are similar (in both form and scale) to those produced by a classical interferometer while also featuring the loss and background resilience inherent in two-photon interference.

\begin{figure}[!htb]
	\includegraphics{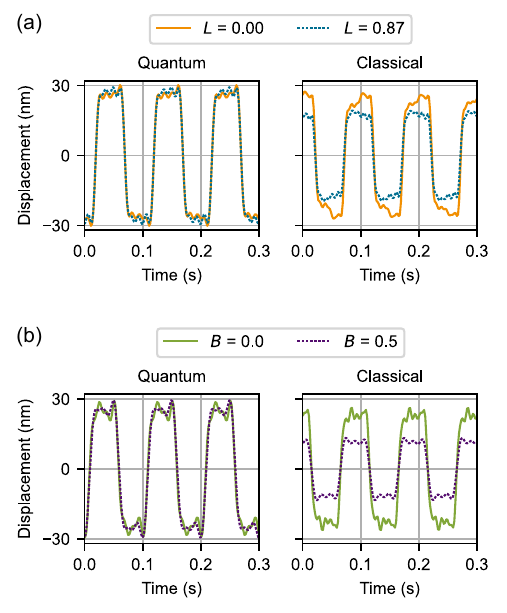}
	\caption{\label{fig:fig_3} Demonstration of quantum advantage. The test signal is a 10-Hz square wave with a true peak-to-peak amplitude of $\sim$55 nm. The quantum measurement is robust against (a) imbalanced path loss ($L$) and (b) optical background ($B$), whereas the classical measurement is not.}
\end{figure}

As discussed in Ref.~\cite{lualdi_fast_2025}, imbalanced path loss acts globally on the two-photon state such that, while the pair-detection rate decreases, the fringe visibility remains unchanged. This contrasts with classical single-photon interference, where imbalanced loss affects the superposition state such that \emph{both} the detection rate and visibility change. Figure \hyperref[fig:fig_3]{\ref*{fig:fig_3}(a)} compares the effect of imbalanced loss for quantum and classical interference when detecting a 10-Hz square wave generated by our nano-positioning stage, with a peak-to-peak amplitude of $\sim$55 nm. We switch between 0\% and 87\% excess loss $L$ in mode $b$ by rotating the half-wave plate upstream of the polarizing beamsplitter (see Fig. \hyperref[fig:fig_1]{\ref*{fig:fig_1}(a)}), and perform classical interference with 1550-nm single photons from our entanglement source. As the classical fringes are also sinusoidal, we modify our analysis protocol by swapping the coincidence detections for single detections in each output of the 50:50 beamsplitter. For analysis, we assume reference fringes from the lossless case.

For the quantum (classical) case, we increased the measurement time after introducing loss from \mbox{$3\rightarrow23$ s} ($1\rightarrow2$ s) to maintain roughly 0.6 million pair (1.2 million single) detections, respectively. Without loss, the quantum (classical) measurement recovered four (six) harmonics and yielded an amplitude of $\sim$60 nm ($\sim$56 nm);  noise in the frequency spectrum precludes recovering higher harmonics. With loss and assuming the lossless reference fringes, the corresponding values are four (five) harmonics and $\sim$59 nm ($\sim$42 nm). We thus see that, unlike frequency estimation, amplitude estimation is severely impacted in the classical case because of loss-induced fringe distortions (visibility and vertical offset). We attribute the slight amplitude overestimate in the quantum case to incomplete harmonic recovery.

We also show background resilience, which is enabled by suppressing accidental detections via our 100-ps coincidence window. These detections decrease the fringe visibility by raising the noise floor; similar temporal filtering in the classical case requires a pulsed light source. We quantify the background $B$ as a fraction of total detected singles, and tune $B$ by shining a halogen lamp onto our apparatus. Figure \hyperref[fig:fig_3]{\ref*{fig:fig_3}(b)} shows how, in the case of our continuous-wave experiment, the quantum measurements are unaffected when increasing $B$ from 0 to 0.5 (corresponding to $\text{SNR} = 1$ for the classical measurement) with total coincidences (over 40 s) remaining at $\sim$300,000; the unused singles increase from $\sim$8 million to $\sim$16 million. In contrast, the classical singles (over 4 s) increase from $\sim$600,000 to $\sim$1.2 million. Since background distorts the classical fringes, the classically estimated amplitude drops from $\sim$52 nm to $\sim$29 nm. The quantum estimation is unaffected, returning $\sim$60 nm for both.

\textit{Conclusions}\textemdash We have demonstrated the sensing of nanometer-scale vibrations with two-photon quantum interference. Our high resolution is made possible by highly non-degenerate energy entanglement, which significantly increases the per-photon information content compared to previous two-photon interferometers. With our bright source of entangled photon pairs, we can thus perform nanometer-scale measurements in a matter of seconds. Such rapid sensing thereby allows us to probe vibrating surfaces and other dynamic systems.

We further enhanced our capabilities by leveraging a ``flux probing'' technique to achieve high-bandwidth sensing without being constrained by the pair-detection rate. Our model is best suited to periodic signals that have significant energy in a relatively sparse set of frequencies. Vibrations with time-varying structure may require a more sophisticated analysis protocol. We also believe our method will still work even as the number of photons detected in each oscillation period becomes very small, although further information-theoretical work is needed to examine the attainable resolution given available temporal and photonic resources.

We experimentally validated our protocol by observing nanometer-scale precision and accuracy for amplitude estimation, along with centi-hertz frequency uncertainty. We also successfully detected vibrational frequencies as high as 21 kHz, which exceeds the nominal upper bound for human hearing (20 kHz) \cite{hansen_noise_2021}. Importantly, we also illustrated the genuine quantum advantage offered by our method for amplitude estimation, which is unaffected by imbalanced loss and optical background. Combined, our results present a new, entanglement-enhanced capability for robust, low-light sensing of time-varying signals at the nanometer scale. 

\vspace{\baselineskip}

\textit{Acknowledgments}\textemdash We thank Pasquale Bottalico, Virginia Tardini, and Charles Nudelman for their acoustics expertise, and Timothy Stelzer for analysis insights. Portions of this Letter are based on a Ph.D. dissertation (C.P.L., \cite{cpl_thesis}). Portions of this work were funded by the U.S. Government. P.G.K., C.P.L., S.J.J., M.V. were partially funded by the U.S. Air Force (FA9550-21-1-0059), P.G.K., C.P.L., S.J.J. were partially funded by the U.S. Department of Energy, Office of Science, Office of Biological and Environmental Research (DE-SC0023167). C.P.L. acknowledges support from the National Science Foundation Graduate Research Fellowship Program (DGE 21-46756). J.R. is exclusively supported by Mitsubishi Electric Research Laboratories. Any opinions, findings, and conclusions or recommendations expressed in this material are those of the authors and do not necessarily reflect the views of the U.S. Air Force, the U.S. Department of Energy, the National Science Foundation, or the U.S. Government.

\vspace{\baselineskip}

\textit{Competing interests}\textemdash S.J.J., P.G.K., and C.P.L. are inventors on U.S. patent no.~12,467,732 (``Precision quantum-interference-based non-local contactless measurement'')

\bibliography{references.bib}

\end{document}